\documentclass[12pt,a4paper]{article}
\usepackage{amsmath}
\usepackage{amsfonts}
\usepackage{graphicx}
\usepackage{psfrag}

\begin{document}
\rightline{DCPT-05/13}   
\rightline{hep-th/0503152}

%% 			Title here  
%%  

\begin{center} 
{\Large \bf Smeared D0 charge and the Gubser-Mitra conjecture}
\end{center} 
\vskip 1cm   
  
\renewcommand{\thefootnote}{\fnsymbol{footnote}}   
\centerline{\bf Simon 
F. Ross\footnote{S.F.Ross@durham.ac.uk}$^{(1)}$ and Toby
Wiseman\footnote{twiseman@fas.harvard.edu}$^{(2)}$}     
\vskip .5cm   
\centerline{ $^{(1)}$ \it Centre for Particle Theory, Department of  
Mathematical Sciences}   
\centerline{\it University of Durham, South Road, Durham DH1 3LE, U.K.}   
\centerline{\it $^{(2)}$ \it Jefferson Physical Laboratory, Harvard
University, Cambridge MA 02138, USA}  

\setcounter{footnote}{0}   
\renewcommand{\thefootnote}{\arabic{footnote}}

%%			Text starts here  
%% 

\begin{abstract}  
We relate a D$p$ or NS-brane with D0-brane charge smeared over its
worldvolume to the system with no D0-charge. This allows us to
generalise Reall's partial proof of the Gubser-Mitra conjecture. We
show explicitly for specific examples that the dynamical instability
coincides with thermodynamic instability in the ensemble where the
D0-brane charge can vary. We also comment on consistency checks of the
conjecture for more complicated systems, using the example of the D4
with F1 and D0 charges smeared on its worldvolume.
\end{abstract}

\section{Introduction}

Black $p$-brane solutions have played a central role in many
investigations in string theory. However, although large classes of
exact solutions have been known for some time, the dynamical features
of the general solutions have not yet been thoroughly explored. Early
on, Gregory and Laflamme showed that uncharged~\cite{gl:unst} and
certain charged~\cite{gl:chunst} $p$-branes are unstable to
perturbations which oscillate in the extended directions of the black
brane with a wavelength longer than the characteristic curvature
scale of the black brane solution. However, these early investigations
were not pursued further until quite recently.

In~\cite{gm1,gm2}, Gubser and Mitra conjectured that the
Gregory-Laflamme instability would occur if and only if the solution
was locally thermodynamically unstable. This conjecture was motivated
by the dual interpretation of the thermodynamics of the black brane in
string theory in terms of a field theory on the brane worldvolume, and
in particular by the fact that the black brane thermodynamics is
extensive in the extended directions. A partial proof for this
conjecture was given by Reall in~\cite{reall}, relating the existence
of a Euclidean negative mode to the mode at the threshold of the
dynamical instability. This argument applied for $p$-brane solutions
carrying magnetic $D-(p+2)$-form charges, encompassing most of the
elementary NS and D-brane solutions in string theory. These arguments
predicted the onset of dynamical instabilities at points where the
specific heat changes sign; numerical evidence for the existence of
these instabilities was obtained in~\cite{cbb1,cbb2,cbb3}.

In this paper, we wish to consider the extension of Reall's
argument~\cite{reall} to study smeared branes; that is, $p$-brane
solutions carrying an electric $q$ form (or magnetic $D-q$ form)
charge where $q < p+2$. We will show that there is a straightforward
extension to the case where the smeared charge is a D0-brane charge
(that is, an electric two-form charge).  Strings carrying a D0-brane
charge were first considered in~\cite{paul}, and we will build on
insights from that work. The case of 2-branes carrying both D2 and
D0-brane charges was considered in~\cite{GubD2D0}, where the dynamical
stability boundary was computed numerically, and shown to agree with
thermodynamic stability. We generalise Reall's arguments and consider
$p$-branes carrying both a magnetic $D-(p+2)$-form charge and the
electric two-form charge, demonstrating the link between thermodynamic
and dynamic stability analytically. We also comment on more
complicated brane systems with smeared D0 charge and exhibit a
thermodynamic consistency condition that must be obeyed for the
Gubser-Mitra relation to hold.

In IIA string theory, we can think of the D0-brane charge in M theory
terms as momentum along the M theory circle. This implies that the M
theory lifts of the solution describing a $p$-brane carrying both a
magnetic $D-(p+2)$-form charge and the D0-brane charge and the
solution without D0-brane charge differ by a boost. Since the
solutions are locally the same in 11 dimensions, we can translate the
argument of~\cite{reall} to a relation between a certain threshold
mode of the solution with magnetic $D-(p+2)$-form charge and D0-brane
charge and a negative mode of the corresponding Euclidean
solution. Thus, we can add D0-brane charge to any case studied in
Reall's argument `for free'.  
Assuming the Gregory-Laflamme instabilities are the only unstable
modes of the dynamical system,
this enables us to predict the stability boundary for these solutions:
in particular, it implies it is independent of the boost parameter
determining the D0-brane charge, as in~\cite{paul}. For the D2-D0
case, the stability boundary we find in this way agrees with that
found in~\cite{GubD2D0}.

One may lift more complicated IIA brane bound states to M theory and
again add smeared D0 charge by boosting the M theory solution. Any
classical instability in the original system will remain in the system
with D0-charge. Hence, if the Gubser-Mitra conjecture is correct, 
the system without D0 charge  has a thermodynamic  
instability independent of the D0-brane charge. 
We explicitly check this for the D4-F1-D0 system and
confirm this holds true for 
thermodynamic ensembles where the D0 brane
charge is allowed to vary.

This use of the M theory perspective develops from a series of
previous insights.  It was observed in~\cite{ho:cyl} that an ansatz
for smeared black holes existed in which the equations were
independent of the black hole's charge under a two-form field
strength. This observation was used in~\cite{paul} to argue that the
dynamical stability of the smeared D0-brane was independent of
charge.\footnote{Dynamical stability for a different class of smeared
charged solutions was studied in~\cite{ls}.}  It was realised
in~\cite{bhbs,ho:nphase} that this simplicity in the ansatz was
attributable to the fact that the charge could be thought of as the M
theory boost, and this observation was exploited in~\cite{bhbs} to
explicitly construct the unstable mode. 

Note that as we find that both the threshold unstable mode and the
Euclidean negative mode can be carried through to the D0-brane charged
case, we are arguing that the Gubser-Mitra conjecture continues to
hold for these cases, contrary to the claims in~\cite{paul}.  As
discussed in these cases in~\cite{bhbs,GubD2D0}, the conjecture is
made consistent provided one chooses a thermodynamic ensemble which
allows the smeared charge to vary.  This is related to the fact that
the smeared charge redistributes over the world-volume in the
dynamical instability.

The structure of the remainder of this paper is fairly simple. In the
next section, we will review the relation between the threshold
unstable mode associated with dynamical instability and the negative
mode associated with thermodynamic instability found in~\cite{reall}
for single $p$-branes. In section~\ref{argument}, we will then
describe how we can change to an M theory description of the
solutions, apply a boost, and reduce to IIA again to obtain a
corresponding relation for solutions which carry an additional D0
brane charge. In section~\ref{eg}, we will discuss specific cases of
D$p$-D0 systems and check the agreement with previous work.  In
section~\ref{complicated} we discuss more general brane bound states
with smeared D0 charge and show that the Gubser-Mitra conjecture again
implies the thermodynamic stability boundary must be independent of
the D0 charge, verifying this for the example of D4-F1-D0.  Finally in
section~\ref{concl}, we conclude with a short discussion of open
questions and future directions.

\section{Connecting thermodynamic and dynamical instability}
\label{review} 

In this section, we review the argument of~\cite{reall} relating
thermodynamic and dynamical instability. Reall considered black
$p$-brane solutions of the action
\begin{equation}
S = \int d^D x \sqrt{-\hat{g}} \left( -\hat{R} + \frac{1}{2} (\partial
\phi)^2 + \frac{1}{2n!} e^{a \phi} F_{(n)}^2 \right),
\end{equation}
where $n=D-(p+2)$, and the $p$-branes were taken to be magnetically
charged under $F_{(n)}$. The cases we will be interested in are the
F1, D2, D4, and NS5-brane solutions in type IIA supergravity (so
$D=10$). For each of these we can reduce the IIA supergravity action
to the above form. For the F1 string, $F_{(7)}$ is the dual of the
Kalb-Ramond 3-form and $a=-1$, for the NS5 brane $F_{(3)}$ is the
Kalb-Ramond 3-form and $a=1$, while for the D2, D4 branes $F_{(6,4)}$
is a RR field strength and $a = (p-3)/2$. 

Later, to consider the extension to include D0-brane charge, we also
include in the action a two-form field strength $F_{(2)}$, so the
total action we consider is
\begin{equation}
S = \int d^10 x \sqrt{-\hat{g}} \left( -\hat{R} + \frac{1}{2} (\partial
\phi)^2 + \frac{1}{2n!} e^{a \phi} F_{(n)}^2 + \frac{1}{4} e^{-\phi/2}
F_{(2)}^2\right).
\end{equation}

In the black $p$-brane solutions, the spacetime coordinates can be
naturally decomposed into coordinates $(t, x^i)$, $i=1,\ldots,p$ in
the directions along the brane, a radial coordinate $r$, and
coordinates $x^m$ on a $8-p$ sphere in the transverse space. The
solutions are invariant under $ISO(p)$ translations and rotations in
the $x^i$ and $SO(9-p)$ rotations on the sphere. We also use a coarser
division of the $D$-dimensional coordinates $x^a$, into the
worldvolume directions $x^i$ and the other directions, collectively
$x^\mu$ (we reserve the notation $x^M$ for 11d coordinates).

To relate dynamical and thermodynamic stability, we are interested in
considering on the one hand the dynamical perturbations of this
solution, and on the other hand, the eigenfunctions of the operator
appearing in the quadratic action for fluctuations around the Euclidean
solution in the Euclidean path-integral approach to the
thermodynamics.

In~\cite{reall}, the perturbation was studied in a different conformal
frame, 
\begin{equation}
g_{ab} =\exp\left( -\frac{(7-p)}{4a} \phi \right) \hat{g}_{ab}.
\end{equation}
The perturbed metric and dilaton are $g_{ab} = \bar{g}_{ab} + h_{ab}$,
$\phi = \bar{\phi} + \delta \phi$. For s-wave perturbations, the
perturbation of the magnetic field strength can be set to zero,
$\delta F_{(n)}=0$.  The perturbation of the metric and dilaton are
assumed to be plane waves in the spatial worldvolume directions,
$h_{ab} = e^{i \mu_i x^i} H_{ab}, \delta \phi = e^{i \mu_i x^i}
f$. Assuming that the perturbation is longitudinal, it can be taken to
be non-zero only in the $x^\mu$ directions by a choice of gauge,
$H_{i\mu} = H_{ij} =0$. 

Our real interest is in dynamical instabilities---that is, in
solutions of the linearised equations of motion which grow
exponentially in time $H_{\mu\nu} \sim e^{\Omega t}$. However, to make
the connection to thermodynamic instability, we will focus on the
threshold unstable mode, which appears at the critical value of
$\mu_i$ such that $\Omega=0$. We will assume that the existence of
such a threshold unstable mode for non-zero $\mu_i$ indicates the
existence of a real instability at longer wavelengths. We thus assume
that the remaining functions $H_{\mu\nu}, f$ are functions only of
$r$. That is, we consider small perturbations that move us in the
space of static solutions, breaking only the symmetry in the spatial
worldvolume directions.

There is then a second-order equation for the dilaton perturbation
$f$, 
\begin{multline} \label{dil}
- \nabla^2 f + 2 \beta \partial_\mu \bar \phi \partial^\mu f +
  H^{\mu\nu} \nabla_\mu \nabla_\nu \bar{\phi} - \beta H^{\mu\nu}
  \partial_\mu \bar \phi \partial_\nu \bar \phi
+ \partial_\mu \bar \phi \nabla_\nu \left(H^{\mu\nu} - \frac{1}{2} H^\rho_\rho
  \bar{g}^{\mu\nu} \right)  \\
- \frac{a}{2 (n-1)!} e^{(\alpha + \beta)
  \bar \phi} \left( H^{\mu\nu} \bar{F}_{\mu \rho_1 \ldots \rho_{n-1}}
  \bar{F}_\nu^{\ \rho_1 \ldots \rho_{n-1}} - \frac{\alpha + \beta}{n}
  \bar{F}^2 f \right) = \lambda f, 
\end{multline}
where $\lambda = -\mu^2$, $\alpha = a-(p-2)(7-p)/2a$, and $\beta =
2(7-p)/a$. The metric perturbations satisfy a second-order equation in
the $x^\mu$ space,
\begin{multline} \label{met}
- \nabla^2 H_{\mu\nu} + 2 \nabla_{(\mu} \nabla^\rho H_{\nu)\rho} -
  \nabla_\mu \nabla_\nu H^\rho_\rho + 2 R_{\rho (\mu} H^\rho_{\ \nu)}
  + 2 R_{(\mu|\rho \sigma|\nu)} H^{\rho\sigma} \\
- \beta \left( 2
  \nabla_{(\mu} H_{\nu)}^{\ \rho} - \nabla^\rho H_{\mu\nu} \right)
  \partial_\rho \bar \phi + 2 \beta \nabla_\mu \nabla_\nu f -
  4(k+\beta^2) \partial_{(\mu} \bar \phi \partial_{\nu)} f \\
+ \frac{1}{(n-1)!} e^{\alpha+\beta} \bar \phi \left( (n-1) H^{\rho
  \sigma} \bar{F}_{\mu\rho \lambda_1 \ldots \lambda_{n-2}}
  \bar{F}_{\nu \sigma}^{\ \ \lambda_1 \ldots \lambda_{n-2}} -
  \bar{F}_{\mu \lambda_1 \ldots \lambda_{n-1}} \bar{F}_\nu^{\
  \lambda_1 \lambda_{n-1}} (\alpha + \beta) f \right) = \lambda
  H_{\mu\nu},
\end{multline}
where $k=1/2-9(7-p)^2/2a^2$, and in addition two constraints coming
from the $\mu i$ and $ij$ components of the metric perturbation
equations,
\begin{equation} \label{con1}
Y_\mu = \nabla_\nu H^\nu_\mu - \beta H^\nu_\mu \partial_\nu \bar{\phi}
 - 2 f (k + \beta^2) \partial_\nu \bar{\phi} = 0,
\end{equation}
\begin{equation} \label{con2}
Z = H^\rho_\rho - 2\beta f =0.
\end{equation}

The above equations define a solution of the linearised perturbation
equations about a given black $p$-brane solution. This perturbation is
independent of $t$ and depends on the spatial worldvolume coordinates
as $e^{i \mu_i x^i}$. The core of the argument of~\cite{reall} was to
show that this could be related to an normalisable off-shell negative
mode of the corresponding Euclidean solution, which is independent of
both $t$ and the $x^i$.

The Euclidean version of the black brane solution is a candidate
saddle-point for the Euclidean path-integral for the canonical
ensemble,
\begin{equation}
Z = \int d[g] d[\phi] d[F] e^{-I[g,\phi,F]},
\end{equation}
where $I$ is the Euclidean action, and the path integral is taken over
all smooth Riemannian geometries with matter fields $\phi$, $F$, with
all fields periodic with period $\beta = T^{-1}$ in imaginary
time. Expanding this action about the saddle point,
\begin{equation} \label{act}
I[g] = I_0[\bar{g},\bar{\phi},\bar{F}] +
I_2[\bar{g},\bar{\phi},\bar{F};\delta g, \delta \phi, \delta F],
\end{equation}
the quadratic part of the action will be of the form 
\begin{equation} \label{qact}
I_2 = \int d^D x \, \delta \psi_\alpha \, \Delta_{\alpha\beta} \,
\delta \psi_\beta, 
\end{equation}
where we use $\delta \psi_\alpha$ to denote collectively the
fluctuations $(\delta g_{ab}, \delta \phi, \delta F_{(n)})$, and
$\Delta_{\alpha\beta}$ is a second-order differential operator
involving the background fields $(\bar{g},\bar{\phi},\bar{F})$. If
there is a normalisable negative mode in the small perturbations
around the solution, that is, if there is a solution of
\begin{equation} \label{evalue}
\Delta_{\alpha\beta} \, \delta \psi_\beta= \lambda \delta \psi_\alpha 
\end{equation}
for $\lambda <0$, then the solution should be interpreted as an
instanton rather than a genuine saddle-point approximation for the
canonical ensemble~\cite{gpy,reall}. 
In~\cite{reall}, it was shown, following~\cite{wy}, that for the black
$p$-brane solutions, negative specific heat implied the existence of a
negative mode, and it was argued that the converse should also be
true.

In~\cite{reall}, it was shown that if we take the ansatz 
\begin{equation}
\delta g_{ab} = H_{ab}, \quad \delta \phi = f, \quad \delta F =0
\end{equation}
for the fluctuation around the saddle-point, then the eigenvalue
equation \eqref{evalue} will reduce precisely to the equations
(\ref{dil},\ref{met}) above. Furthermore, (\ref{con1},\ref{con2}) are
necessary conditions for the negative mode to be normalisable; they
correspond to an appropriate choice of gauge in which the eigenvalue
equation can really be written in the form
(\ref{dil},\ref{met}). Thus, the conditions for the existence of a
normalisable negative mode are precisely the same set of equations
(\ref{dil}-\ref{con2}) which determine a threshold unstable mode.  The
two problems were also shown to involve the same boundary
conditions~\cite{reall}.  Since $\delta F = 0$ for these static linear
modes, and they reduce to negative modes of the Euclidean action, the
instability is at fixed $n$-form charge. The appropriate ensemble
thermodynamically is then the canonical one, namely fixing the
$p$-brane charge.

\section{Adding D0-brane charge}
\label{argument}

We now want to consider adding D0-brane charge to the black $p$-brane
solutions. Here, we will exploit the fact that these IIA supergravity
solutions can be simply related to M theory; if we have a IIA
supergravity solution with metric $g_{ab}$, dilaton $\phi$ and
$n$-form field strength $F_{(n)}$, $n=3,4,6,7$, this can be used to
construct a solution of M theory with 
\begin{equation}
ds^2_{(11)} = e^{4/3 \phi} dz^2 + e^{-2/3 \phi} ds^2_{(10)},
\end{equation}
and a four-form field strength $H_{(4)}$ given by 
\begin{equation}
H_{(4)} = \left\{ \begin{array}{c} F_{(3)} \wedge dz, \\ F_{(4)}, \\ \star
    (F_{(6)} \wedge dz), \\ \star F_{(7)} \end{array} \right.
\end{equation}
respectively. The 11d solution corresponding to the original IIA
supergravity solution has a compact $z$ direction, but let us consider
the non-compact solution. We can then boost it in the $t-z$ plane,
defining coordinates
\begin{equation} \label{boost}
\begin{aligned}
t' &= t \cosh \beta + z \sinh \beta, \\
z' &= z \cosh \beta + t \sinh \beta.  
\end{aligned}
\end{equation}
If we then compactify the $z'$ direction, we can define a new IIA
background by
\begin{equation}
ds^2_{(11)} = e^{4/3 \phi'} (dz' + A)^2 + e^{-2/3 \phi'} ds^{'2}_{(10)},
\end{equation}
which will carry a D0-brane charge under the two-form field strength
$F_{(2)} = dA$. We will refer to this combined operation as a twist.

Starting with a black $p$-brane solution, this operation will give a
D0-brane charge which is smeared over all the spatial worldvolume
directions, as the $ISO(p)$ symmetry in these directions is
preserved. The threshold unstable mode of the original black $p$-brane
geometry discussed in the previous section corresponds to going a
small distance along a new branch of solutions which are not
translationally invariant in the spatial worldvolume directions. We
can apply the above boost + recompactification to this branch of
solutions; hence, there is a corresponding threshold unstable mode for
the resulting black $p$-brane with a smeared D0-brane charge. For the
special case of an uncharged black string, this charge independence of
the threshold mode was observed in~\cite{paul} using the observations
of~\cite{ho:cyl}. This twist argument provides a general understanding
of that result (as previously observed in~\cite{ho:nphase,bhbs}). 

If we take the ansatz given in the previous section and use this twist
to translate it into a threshold unstable mode for the black $p$-brane
with non-zero D0-brane charge, we will find that the non-zero
components of the metric perturbation are still of the form $\delta
g_{ab} = e^{i \mu_i x^i} H'_{ab}$, and $\delta \phi = e^{i \mu_i x^i}
f'$, but starting from the ansatz given above, we will have $H'_{ij}$
non-zero. We will also have a perturbation $\delta A = e^{i \mu_i x^i}
a$, but we still have $\delta F_{(n)} = 0$. The expressions for
$H_{ab}'$, $f'$ and $a$ in terms of $H_{\mu\nu}$ and $f$ depend on the
background $\bar{g}_{\mu\nu}$ and $\bar{\phi}$, but are
straightforward to derive; we will not give them explicitly. Note that
the critical wavelength will still have the same value $\mu$,
independent of the D0-brane charge parameter $\beta$. Thus, the
existence of this threshold unstable mode is independent of the
D0-brane charge for any of the black $p$-brane solutions with smeared
D0-brane charge.\footnote{The $p$-branes which we can treat by this
argument are the F1, D2, D4 and NS5-branes in IIA. The D6 is not
included in this discussion, as it corresponds to a KK monopole in M
theory, so we cannot take the $z$ coordinate non-compact in the 11d
solution. Recall also that the existence and critical wavelength of
the instability will depend on the $p$-brane charge.}

The fact that the critical wavelength is independent of $\beta$ is
somewhat surprising. We expect that the BPS smeared D0-brane solution
will be marginally stable, but this result indicates that there will
be an instability at wavelengths longer than $\mu$ as soon as we move
away from extremality. In~\cite{bhbs}, these instabilities were
studied explicitly, and it was found that they grow in time as
$e^{\tilde \Omega t}$, where $\tilde \Omega \propto 1/\cosh \beta$, so
$\tilde \Omega \to 0$ as $\beta \to \infty$, consistent with the
expected stability of the BPS solution. 

We should now consider if there is a corresponding Euclidean negative
mode. The Euclidean negative mode of the black $p$-brane solution will
define a corresponding $z$-independent negative mode of the M theory
solution. The IIA supergravity action in (\ref{act},\ref{qact}) can be
thought of as a consistent truncation of the full 11d supergravity
action to a $z$-independent sector, so any eigenvector of \eqref{evalue}
will also be an eigenvector of the corresponding operator in the full
theory.

One might think that the Euclidean analogue of \eqref{boost} would be
a rotation in the $\tau-z$ plane; that is, in addition to the analytic
continuation $t \to i\tau$, we might imagine also making the analytic
continuation $\beta \to i\gamma$. However, recall that our desire here
is to use the Euclidean path integral as a tool to elucidate the
thermodynamic properties of a Lorentzian solution carrying electric
(D0-brane) charge. It was argued in~\cite{medual} that to describe the
thermodynamics of electrically charged solutions, one should not make
an analytic continuation of the electric field; one should accept that
the electric field will be imaginary in the Euclidean
section. In~\cite{medual} this was important to maintaining invariance
under electric-magnetic duality; the usual Hodge duality in Euclidean
space will relate real magnetic to imaginary electric fields. In the
same way, here, we will interpret the Euclidean version of the black
$p$-brane with D0-brane charge obtained by analytically continuing $t
\to i\tau$, but keeping $\beta$ real as the appropriate saddle-point
for the path integral in the canonical ensemble.\footnote{One can
avoid discussing imaginary fields by regarding the D0-brane charge as
a magnetic charge under the dual 8-form field strength. The threshold
unstable mode can also be dualised to rewrite it as a threshold
unstable mode of the magnetic solution, which will be related to a
real Euclidean negative mode on the Euclidean saddle-point
solution. However, this dualising conceals the 11d origin of the
simple behaviour of the instability.} From the 11d point of view, this
saddle point has a complex metric; this is similar to the discussion
of rotating black hole solutions in~\cite{bycx}.

Since the redefinition \eqref{boost} is just a coordinate
transformation, albeit now a complex one, it does not change the fact
that the solution of \eqref{evalue} is a negative mode. Since the
original solution was independent of both $t$ and $z$, this will be
$z'$-independent, and we can reduce it to obtain a negative mode
around the saddle-point constructed from the Euclidean D$p$-smeared D0
solution in the IIA theory. The explicit form of this negative mode is
again given by taking the threshold unstable mode, dropping the
dependence on the spatial worldvolume coordinates, and analytically
continuing $t \to i\tau$. Thus, the existence of the negative mode is
also independent of the D0-brane charge, and these two instabilities
remain correlated.

As before, $\delta F_{(n)}=0$, but we now have $\delta A \ne
0$. Following~\cite{reall}, it seems natural to interpret this
Euclidean negative mode as indicating an instability in the ensemble
where we fix the $n$-form charge, but allow the D0-brane charge to
vary, as in~\cite{bhbs,GubD2D0}. This is also in accord with the
original argument of~\cite{gm1,gm2}, who concluded that the dynamical
instability should be linked to thermodynamic instability in the
ensemble where all extensive parameters are allowed to vary. We will
not attempt to generalise the argument of~\cite{wy,reall} to show that
thermodynamic instability in this sense implies the existence of a
negative mode in this more general case. Instead, we will now turn to
the consideration of specific examples, and verify that in all cases
the region of thermodynamic instability in this ensemble coincides
with the region where this negative mode exists.

\subsection{Examples}
\label{eg}

We would like to now illustrate this general discussion with some
specific examples, and explicitly demonstrate the connection between
the dynamical instability and thermodynamic instability.  The simplest
case to consider is when we start with an uncharged black $p$-brane
solution, so the twist gives us just a non-extremal black D0-brane
smeared over $p$ transverse directions. This case was considered
in~\cite{paul,bhbs}, where it was observed that the dynamical
instability is independent of the charge. The metric in this case is
(we give solutions in Einstein frame)
\begin{equation}
ds^2 = - H^{-7/8} f dt^2 + H^{1/8} ( f^{-1} dr^2 + r^2 d\Omega_{8-p}^2 +
d\vec{x}^2), 
\end{equation}
and the one-form gauge potential and dilaton are  
\begin{equation}
A = \coth \beta \left( 1 - H^{-1} \right) dt, \quad e^{2\phi} = H^{3/2},
\end{equation}
where
\begin{equation}
H = 1 + \frac{r_0^{7-p} \sinh^2 \beta}{r^{7-p}}, \quad f = 1 -
\frac{r_0^{7-p}}{r^{7-p}}.
\end{equation}
We can read off the mass, charge and thermodynamic parameters from
this solution,
\begin{align}
M &= \frac{\Omega_{8-p} V_p}{16 \pi G} r_0^{7-p} ((8-p) + (7-p) \sinh^2
\beta ), \\
Q_0 &= \frac{\Omega_{8-p} V_p}{16 \pi G} (7-p) r_0^{7-p} \sinh \beta
\cosh \beta, \\
S &= \frac{\Omega_{8-p} V_p}{4 G} r_0^{8-p} \cosh \beta, \quad T =
\frac{(7-p)}{4\pi r_0 \cosh \beta}, \quad \Phi_0 = \tanh \beta. 
\end{align}
In~\cite{paul}, the charge was regarded as fixed, and the
thermodynamic stability was studied by considering the specific heat,
$C_Q = \left( \frac{\partial M}{\partial T} \right)$. This changes
sign near (but a finite distance from) extremality for $p<5$, which
was interpreted as an indication of thermodynamic stability, violating
the Gubser-Mitra conjecture.

However, as we have discussed in the previous section, we should
consider the thermodynamic ensemble which allows the D0 charge to
fluctuate.  That is, we should also consider the isothermal
permittivity, $\epsilon_T = \left( \frac{\partial Q}{\partial \Phi}
\right)_T$, which probes the thermodynamic stability under changes of
the charge. This is negative, indicating an instability, in precisely
the region where the specific heat becomes positive, so there is
indeed a thermodynamic instability for all values of the
charges. Explicitly~\cite{cbb1},
\begin{align}
C_Q &= - \frac{\Omega_{8-p} V_p}{4 G} r_0^{8-p} \cosh \beta
  \frac{(7-p)+(9-p) \cosh 
  2\beta}{(7-p)-(5-p) \cosh 2\beta} \\
\epsilon_T &= \frac{(7-p) \Omega_{8-p} V_p}{32 \pi G} r_0^{7-p}
  \cosh^2 \beta \left( (7-p)-(5-p) \cosh 2\beta \right), 
\end{align}
so if $p<5$, $C_Q >0$ for sufficiently large $\beta$, but we never
have $C_Q >0$ and $\epsilon_T >0$, indicating that the smeared
D0-brane solution is always unstable in the grand canonical ensemble,
as expected from the existence of the Euclidean negative mode
obtained by our general argument.  

It is interesting to note that these smeared D0-brane solutions are
also related to the charged D$p$-branes discussed in~\cite{reall} by
T-duality in the spatial worldvolume directions, but this relation is
not helpful for understanding the stability. The dynamical instability
depends on the worldvolume directions, so it would not be expected to
be invariant.  Rather these unstable gravity modes should presumably
be thought of as unstable stringy winding modes in the T-dual theory.
The thermodynamic stability fails to be invariant for a subtler
reason: the thermodynamic quantities are invariant under T-duality,
but the appropriate ensemble is the canonical one (fixed charge) for
the D$p$-brane solution, but the grand canonical one (charge allowed
to vary) for the smeared D0-brane solution~\cite{bhbs}.

The smeared D0-brane case is thus seen to be consistent with the
Gubser-Mitra conjecture, once we understand the proper application of
the conjecture in this case. It is more interesting to consider a case
with a non-trivial boundary of stability, which will provide a sharp
test of the conjecture. The case of D2-branes with smeared D0-branes
was studied in~\cite{GubD2D0}, where a numerical analysis of the
Lorentzian linear dynamics indicated consistency of the Gubser-Mitra
conjecture. We may now show this analytically.  We start from a
solution with D2-brane charge,
\begin{equation}
ds^2 = Z^{-5/8} \left[ -fdt^2 + dx_1^2 + dx_2^2 \right] +
Z^{3/8}  ( f^{-1} dr^2 + r^2 d\Omega_6^2),
\end{equation}
with gauge field and dilaton
\begin{equation}
F_{(6)} = 5 r_0^5 \sinh \gamma \cosh \gamma \epsilon_{S^6}, \quad
e^{2\phi} = H^{1/2},
\end{equation}
where
\begin{equation} \label{fZ}
Z = 1 + \frac{r_0^5 \sinh^2 \gamma}{r^5}, \quad f = 1 -
\frac{r_0^5}{r^5},
\end{equation}
and $\epsilon_{S^6}$ is the volume form on the unit six-sphere,
corresponding to the metric $d\Omega_6^2$. Following~\cite{reall} and
our general discussion above, we are writing the D2-brane charge as a
magnetic charge under $F_{(6)}$. We can obtain a solution with D2-brane
and smeared D0-brane charge by applying the twist. This gives a
solution
\begin{equation}
ds^2 = H^{-5/8} \left[ -fdt^2 + H Z^{-1} (dx_1^2 + dx_2^2 \right)] +
H^{3/8}  ( f^{-1} dr^2 + r^2 d\Omega_6^2),
\end{equation}
with a one-form gauge field and dilaton
\begin{equation}
A = \frac{\cosh^2 \gamma \sinh \beta \cosh \beta}{ \sinh^2 \gamma
  \cosh^2 \beta + \sinh^2 \beta} \left( 1 - H^{-1} \right) dt, \quad
e^{2\phi} = H^{3/2} Z^{-1},
\end{equation}
and the D2-brane gauge field is unchanged, 
\begin{equation}
F_{(6)} = 5 r_0^5 \sinh \gamma \cosh \gamma \epsilon_{S^6},
\end{equation}
where
\begin{equation}
H = Z \cosh^2 \beta - f \sinh^2 \beta = 1 + \frac{r_0^5 (\sinh^2
  \gamma \cosh^2 \beta + \sinh^2 \beta)}{r^5}, 
\end{equation}
and $f$, $Z$ are as in \eqref{fZ}. This solution was written in a
different form by Gubser in~\cite{GubD2D0}, where the D2-brane charge
was written in terms of a gauge field $A_3$; the twist then produces a
non-zero NS two-form $B_2$. He also uses a function $D = H Z^{-1}$, and
parameters $\alpha$, $\theta$ related to the ones used here by
\begin{equation}
\sinh^2 \alpha = \sinh^2
  \gamma \cosh^2 \beta + \sinh^2 \beta
\end{equation}
and
\begin{equation}
\sin^2 \theta = \frac{\cosh^2 \gamma \sinh^2 \beta}{\sinh^2
  \gamma \cosh^2 \beta + \sinh^2 \beta}. 
\end{equation}
This implies
\begin{equation}
\sinh^2 \gamma = \sinh^2 \alpha \cos^2 \theta. 
\end{equation}

Our general twist argument predicts that the instability will be
independent of $\beta$: this is in agreement with the results of
Gubser~\cite{GubD2D0}, who finds a thermodynamic stability boundary at
\begin{equation}
\sinh \gamma = \sinh \alpha \cos \theta = \frac{1}{\sqrt{3}}, 
\end{equation}
which is confirmed by numerical studies of the dynamical
instability. In this case, the thermodynamic instability is at fixed
D2-brane charge, but allowing the D0-brane charge to vary,
corresponding to the fact that $\delta F_{(6)} = 0$ but $\delta
A \neq 0$ in the ansatz for the threshold unstable mode and the
negative mode. 

Another interesting case to consider is a D4-brane with smeared
D0-brane charge. The solution is
\begin{equation}
ds^2 = H_0^{-7/8} H_4^{-3/8} (-f dt^2 + H_0 d\vec{x}^2 ) + H_0^{1/8} H_4^{5/8}
(f^{-1} dr^2 + r^2 d\Omega_4^2),
\end{equation}
with gauge fields and dilaton
\begin{equation}
A = \coth \beta \left( 1 - H_0^{-1} \right) dt, \quad F_{(4)} = 3
r_0^3 \sinh \alpha \cosh \alpha \epsilon_{S^4}, \quad e^{2\phi} =
H_0^{3/2} H_4^{-1/2},
\end{equation}
where
\begin{equation}
H_0 = 1 + \frac{ r_0^3 \sinh^2 \beta}{r^3}, \quad H_4 = 1 + \frac{
  r_0^3 \sinh^2 \alpha}{r^3}, \quad f = 1 - \frac{r_0^3}{r^3},
\end{equation}
and $\epsilon_{S^4}$ is the volume form on the unit four-sphere.  The
parameter dependence in this case is simpler: from the 11d point of
view, the D4-brane comes from a reduction along a worldvolume
direction in the M5-brane, whereas to obtain a D2-brane we had to
smear the M2-brane over a transverse direction. From the 10d point of
view, the simplicity is related to the fact that the BPS D4-D0 is a
marginally bound system. The thermodynamic parameters are
\begin{align}
M &= \frac{\Omega_4 V_4}{16 \pi G} r_0^3 [ 4 + 3 \sinh^2 \alpha + 3
  \sinh^2 \beta], \\
Q_4 &= \frac{\Omega_4}{16 \pi G} 3 r_0^3 \sinh \alpha \cosh \alpha, \\
Q_0 &= \frac{\Omega_4 V_4}{16 \pi G} 3 r_0^3 \sinh \beta \cosh \beta, \\
S &= \frac{\Omega_4 V_4}{4 G} r_0^4 \cosh \alpha \cosh \beta,\quad T
  = \frac{3}{4\pi r_0 \cosh \alpha \cosh \beta}, \\
\Phi_0 &= \tanh
  \beta, \quad \Phi_4 = \tanh \alpha. 
\end{align}
We again want to calculate the Hessian of the entropy with respect to
$M$ and $Q_0$, holding $Q_4$ fixed. Evaluating this Hessian, we find
\begin{equation}
\det H = \frac{(\frac{4}{9} \pi)^2 (16 \pi G)^2 (\sinh^4 \alpha -1)}
  {\Omega_4^2 V_4^2 
  r_0^4 (4 \sinh^2 \alpha \sinh^2 \beta + 5 \sinh^2 \alpha + 5 \sinh^2
  \beta + 4) }. 
\end{equation}
The boundary of thermodynamic stability is at $\det H =0$, so we see
that it is independent of $\beta$, as expected from our general
argument. The system is unstable for $\sinh^2 \alpha <1$ and stable
for $\sinh^2 \alpha >1$. In figure 1, we plot this stability boundary
in the $Q_4$, $Q_0$ plane for fixed $M$. We plot the square roots of
the charges to give the BPS bound a similar form to the D2-D0 case
studied in~\cite{GubD2D0}.

\begin{figure}[htbp]
\centering
\psfrag{Q0}{\small $\sqrt{Q_0/M}$}
\psfrag{Q4}{\small $\sqrt{Q_4 V_4/M}$}
\includegraphics[width=0.6\textwidth]{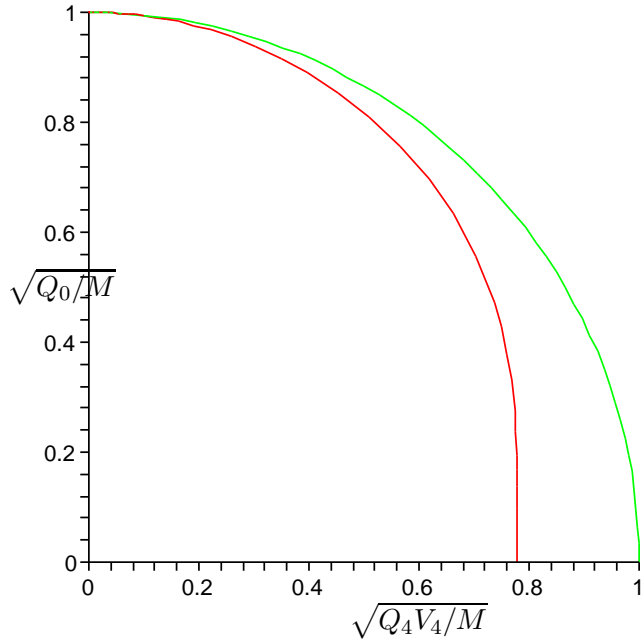}
\caption{The boundary of stability for the D4-smeared D0 system, as a
function of $\sqrt{Q_4 V_4/M}$ and $\sqrt{Q_0/M}$. The system is
stable to the right of the line. The circle indicates the BPS bound.}
\label{fig1}
\end{figure}

We can also consider F1 and smeared D0 or NS5 and smeared D0. We will
not describe these two cases in detail, as they add no really new
elements. The F1-D0 case is qualitatively similar to the D4-D0 case
discussed above, while the NS5-D0 is not as interesting, as there is
no stability boundary: the system is always both thermodynamically and
dynamically unstable.

\section{Adding D0 charge to general brane bound states}
\label{complicated}

Provided a translationally invariant brane bound state can be lifted
to M theory, much of the discussion above still applies. Take a IIA
system $X$, lift to M theory and boost. Reducing then gives a smeared
$X$-D0 system.  For example D4-F1, with the F1 charge smeared on the
D4, lifts to M5-M2 and after a twisted reduction gives the D4-F1-D0
system.

The principal difference from our previous discussion is that for
general systems $X$ we have no argument to relate classical and
thermodynamical stability for $X$.  However, what remains true is that
any classical instability of $X$ will lift to an instability in M
theory. The twisted reduction boosts on the 11-circle and thus does
not effect this instability mode. Hence the $X$-D0 system will exhibit
the same instability for any D0 charge; as before, the critical
wavelength for the instability will be independent of the D0 brane
charge parameter $\beta$, but the instability will become weaker as we
approach extremality.

This provides a consistency check of the Gubser-Mitra conjecture. If
it holds for both $X$ and $X$-D0, then the classical stability of $X$
and $X$-D0 will be connected to the thermodynamic stability of these
systems. This implies that the thermodynamic instability boundary for
$X$-D0 must be independent of the D0 charge parameter, where we consider
thermodynamic stability with the D0 charge allowed to vary.

Consider for example D4-F1-D0. 
The metric for the D4-F1 is found in~\cite{harmark} and using the
twisted reduction method we may add D0-brane charge. Doing so gives
thermodynamic relations similar in form to those of D4-D0 considered
above, except there is an additional parameter $\theta$ related to the
presence of the F1's,
\begin{align}
M &= \frac{\Omega_4 V_4}{16 \pi G} r_0^3 [ 4 + 3 \sinh^2 \alpha + 3
  \sinh^2 \beta], \\
Q_4 &= \frac{\Omega_4}{16 \pi G} 3 r_0^3 \sinh \alpha \cosh \alpha
  \cos \theta, \\
%% SFR \beta -> \alpha
%Q_1 &= \frac{\Omega_4 V_4}{16 \pi G} 3 r_0^3 \sinh \beta \cosh \beta \sin \theta, \\
Q_1 &= \frac{\Omega_4 V_4}{16 \pi G} 3 r_0^3 \sinh \alpha \cosh \alpha \sin \theta, \\
Q_0 &= \frac{\Omega_4 V_4}{16 \pi G} 3 r_0^3 \sinh \beta \cosh \beta, \\
S &= \frac{\Omega_4 V_4}{4 G} r_0^4 \cosh \alpha \cosh \beta,\quad T
  = \frac{3}{4\pi r_0 \cosh \alpha \cosh \beta}, \\
\Phi_0 &= \tanh
  \beta,\quad \Phi_1 = \tanh \alpha \sin \theta, \quad \Phi_4 = \tanh \alpha \cos \theta. 
\end{align}
In considering the Gubser-Mitra conjecture, the thermodynamic ensemble
whose stability is related to dynamic stability will have the D4
charge fixed.  For the F1 charge, the same logic suggests that the
ensemble where the F1 charge is fixed will be related to dynamic
stability against perturbations which vary only in the directions
along the F1 string, while the ensemble where the F1 charge is allowed
to vary will be related to perturbations which vary in the directions
perpendicular to the F1 string in the D4 worldvolume. We will consider
both ensembles, firstly with fixed F1 charge and secondly with it
varying.

For the first ensemble we consider the Hessian of the entropy with
respect to $M$ and $Q_0$. This simply gives the same answer as for the
D4-D0 system of the previous section. Hence the stability boundary is
indeed independent of $\beta$, the D0 charge parameter.

For the second ensemble, where we allow the F1 and D0 charge to
vary, the stability is given by the Hessian of the entropy with
respect to $M$, $Q_0$ and $Q_1$, holding just $Q_4$ fixed. One finds
\begin{equation}
\det H = \frac{(\frac{4}{9} \pi)^3 (16 \pi G)^3 \cosh \alpha \cosh \beta (\sinh^2 \alpha \cos^2 \theta - 1 )}
  {\Omega_4^3 V_4^3 
  r_0^6 (4 \sinh^2 \alpha \sinh^2 \beta + 5 \sinh^2 \alpha + 5 \sinh^2
  \beta + 4) }. 
\end{equation}
and hence the stability region is given simply by $\sinh^2 \alpha
\cos^2 \theta >1$, and again passes our consistency check being
independent of the D0 charge parameter $\beta$. 

Note that we have parametrized the D4 and F1 charges in terms of
$\alpha,\theta$, in the same way that Gubser~\cite{GubD2D0}
parametrized the D2-D0 system. This is convenient 
for studying the stability in the ensemble where we fix both the D4
and F1 charges. However, in the ensemble where we allow the F1 charge
to vary, it would be more natural to use a description analogous to
the one we used in the previous section for D2-D0; that is, to have
independent D4 and F1 charge parameters $\gamma, \delta$ such that
\begin{equation}
\sinh^2 \alpha = \sinh^2
  \gamma \cosh^2 \delta + \sinh^2 \delta
\end{equation}
and
\begin{equation}
\sin^2 \theta = \frac{\cosh^2 \gamma \sinh^2 \delta}{\sinh^2
  \gamma \cosh^2 \delta + \sinh^2 \delta}. 
\end{equation}
Then we have
\begin{equation}
\sinh^2 \gamma = \sinh^2 \alpha \cos^2 \theta, 
\end{equation}
and the stability boundary is independent of the F1 charge parameter
$\delta$, as in the discussion of the D2-D0 case. 

This independence can be understood by relating this system to a
system with smeared D0 branes by U-duality. From D2-D0, a T-duality in
a transverse direction followed by S-duality followed by another
T-duality in a transverse direction will give us the D4-F1
solution. Since these T-dualities require us to smear the system over
additional transverse directions, the D4-F1 system is really U-dual to
a system with D2-branes smeared over two transverse directions and
D0-branes smeared over all four directions. The stability of this
system thus need not be related to that of the D2-D0 system we studied
in the previous section, and in fact, the stability boundary has
changed: it's at $\sinh^2 \gamma =1$ for D4-F1, but it was at $\sinh^2
\gamma = 1/3$ for D2-D0. Nonetheless, we can regard the D2-brane
smeared over two transverse directions as another system $X$ to which
we are adding smeared D0-brane charge by twisting. The threshold
unstable modes of the D4-F1 system are related to the threshold
unstable modes of the smeared D2-D0 system which only depend on the
directions along the D2-brane's worldvolume, so these threshold
unstable modes should be independent of the charge parameter
$\delta$. As in the argument we gave earlier in this section, this
then predicts that the thermodynamic stability boundary will be
independent of the F1 charge parameter $\delta$.

\section{Conclusions}
\label{concl}

We have argued that the Gubser-Mitra conjecture holds for a number of
$p$-brane systems with smeared D0-branes, by extending the argument
given in~\cite{reall}.  These arguments explicitly show the classical
stability agrees with the Euclidean negative mode.  We also showed in
a number of examples that this agrees with the boundary of local
thermodynamic stability of the ensemble where the D0-charge is allowed
to vary, again agreeing with Gubser \& Mitra's idea that all extensive
worldvolume quantities should be allowed to fluctuate.  Furthermore,
we have found that in those cases where there is a boundary separating
unstable from stable solutions (namely D2-D0, D4-D0 and F1-D0), this
boundary takes a simple form: it is independent of the parameter
$\beta$ associated with the D0-brane charge. These results were
obtained by relating the solutions with D0-brane charge to the plain
$p$-brane solutions discussed in~\cite{reall} via an 11d M theory
solution, where the parameter $\beta$ is interpreted as parameterising
a boost in the additional direction.

We have also argued that if the Gubser-Mitra conjecture is correct,
more complicated brane systems with smeared D0 charge must have
thermodynamic instability boundaries independent of this D0 charge,
since we have shown the dynamical instabilies are independent of
this. This provides a consistency check of the conjecture. 
We have verified that it is satisfied in the D4-F1-D0 system.

This discussion seems to be very special to the case of D0-branes, and
it is not clear how to extend it to other cases. However, the
resulting relationship between the threshold unstable mode and the
negative mode for the solutions with smeared D0-brane charge has much
the same flavour as in the case without smeared charge. The negative
mode is just obtained by dropping the dependence on the spatial
worldvolume coordinates in the threshold unstable mode and
analytically continuing to the Euclidean solution, $t \to i \tau$. It
is the complicated form for the ansatz for the threshold unstable mode
allowing us to reduce the equations of motion to a form independent of
$\beta$ that is the key insight obtained from the twist.

We can also obtain some further results for other cases by
T-duality. We have already noted that the T-duality relating the
D0-brane smeared over $p$ directions to the D$p$-brane is not
interesting for understanding the instability, as the instability
depends on these directions. However, if we consider a D0-brane
smeared over $p+q$ directions for $q>0$, we can consider an
instability which is oriented so it only depends on $q$ of the spatial
coordinates. We can then T-dualise in the other $p$ directions to
obtain a D$p$-brane smeared over $q$ transverse directions. The
instability will be invariant under this transformation, and the grand
canonical ensemble will still be the relevant one for the smeared
D$p$-brane. We can therefore use our results above to conclude that
any D$p$-brane smeared over transverse directions will always be both
dynamically and thermodynamically unstable, and that the wavelength of
the dynamical instability will be independent of the charge parameter
$\beta$.

Similarly, we can T-dualise the D2-D0 system on one direction to
obtain an orthogonally intersecting pair of D1's smeared over the two
worldvolume directions, and we can T-dualise the D4-D0 over one or two
directions to obtain orthogonal D3-D1 and D2-D2 systems. These will
all have the same stability properties as the cases studied here.

\medskip
\centerline{\bf Acknowledgements}
\medskip    

The work of SFR is supported by the EPSRC. TW was supported by the
David and Lucille Packard Foundation, grant number 2000-13869A. We
thank Troels Harmark for pointing out an error in the previous version
of the paper.

\bibliographystyle{utphys}  
 
\bibliography{smeared}   

\end{document}